\documentclass[twocolumn,aps,showpacs,showkeys]{revtex4}
\usepackage[T1]{fontenc}
\usepackage[latin1]{inputenc}
\usepackage{amsmath}
\usepackage{graphicx}
\usepackage{amssymb}

\newcommand{\noun}[1]{\textsc{#1}}
\newcommand{\Ain}{A_{\text{in}}}
\newcommand{\Aout}{A_{\text{out}}}
\newcommand{\Vex}{V_{\text{ex}}}
\newcommand{\theq}{\theta_{\text{eq}}}
\newcommand{\Hfilm}{H_{\text{film}}}
\newcommand{\Real}{\text{Re}\,}

\begin{document}

\title{Stability of liquid ridges on chemical micro- and nanostripes}

\author{S. Mechkov}

\affiliation{Laboratoire de Physique Th\'{e}orique de la Mati\`{e}re Condens\'{e}e, Universit\'{e}
Pierre et Marie Curie, Tour 24, Bo\^{\i}te 121, 4 place Jussieu, F-75252
Paris 05, France}

\affiliation{Laboratoire de Physique Statistique, Ecole Normale Sup\'{e}rieure, 24
rue Lhomond, F-75231 Paris Cedex 05, France}

\author{M. Rauscher}

\email{rauscher@mf.mpg.de}

\affiliation{Max-Planck-Institut f\"{u}r Metallforschung, Heisenbergstr.\,3, D-70569
Stuttgart, Germany}

\affiliation{Institut f\"{u}r Theoretische und Angewandte Physik, Universit\"{a}t Stuttgart,
Pfaffenwaldring 57, D-70569 Stuttgart, Germany}

\author{S. Dietrich}

\affiliation{Max-Planck-Institut f\"{u}r Metallforschung, Heisenbergstr.\,3, D-70569
Stuttgart, Germany}

\affiliation{Institut f\"{u}r Theoretische und Angewandte Physik, Universit\"{a}t Stuttgart,
Pfaffenwaldring 57, D-70569 Stuttgart, Germany}

\date{\today}

\begin{abstract}
We analyze the stability of sessile filaments (ridges) of nonvolatile
liquids versus pearling in the case of externally driven flow along
a chemical stripe within the framework of the thin film approximation.
The ridges can be stable with respect to pearling even if the contact
line is not completely pinned. A generalized stability criterion for
moving contact lines is provided. For large wavelengths and no drive,
within perturbation theory, an analytical expression of the growth
rate of pearling instabilities is derived. A numerical analysis shows
that drive further stabilizes the ridge by reducing the growth
rate of unstable perturbations, even though there is no complete stabilization.
Hence the stability criteria established without drive ensure overall
stability.
\end{abstract}

\pacs{68.15.+e, 68.03.Cd, 68.03.-g}

\keywords{pearling instability, chemical channel, microfluidics}

\maketitle

\section{Introduction}

In the last decade substantial efforts have been invested in integrating
chemical processes into microfluidic systems known as ``labs on
a chip'' \cite{stone01,mitchell01,giordano01}. These microfluidic
devices do not only allow for cheap mass production but they can also
operate with much smaller quantities of reactants and reaction products
than standard laboratory equipments.

In this context both closed and open channel systems are considered
for fluid transport. While closed channels are prone to clogging by,
e.g., colloids or large biopolymers, the fluid in open channel systems
has less friction because it is in contact with less substrate material,
and production is possibly cheaper. The substrate surfaces can be
structured chemically by printing or photographic techniques. Conceptually,
the liquid is guided by lyophilic stripes on an otherwise lyophobic
substrate \cite{gau99,darhuber01,dietrich05,zhao02}, i.e., it is
confined by laterally varying substrate potentials, acting as ``chemical
walls''.

Here we analyze the stability of homogeneously filled chemical channels
with respect to pearling, i.e., breakup into a string of droplets.
For all contact angles, on \emph{homogeneous} substrates sessile filaments
(ridges) of nonvolatile liquids are unstable with respect to pearling,
even in the presence of line tension \cite{brinkmann02,brinkmann05,mechkov07}.
However, in the cases that the contact line is infinitely stiff or
pinned, e.g., at the edges of a chemical channel formed by a lyophilic
stripe on an otherwise lyophobic substrate, the instability is suppressed
if the contact angle of the liquid-vapor interface with the substrate
is smaller than $90^{\circ}$ \cite{davis80,brinkmann04}. Molecular
dynamics simulations have confirmed this even at the nano-scale \cite{koplik06a}.

From these results it is not clear whether complete pinning is required
to stabilize a liquid ridge. In an actual sample the three phase contact
angle will not vary step-like at the channel edge. One rather expects
a gradual transition, which leads to a \emph{partial} stabilization:
the contact line can move but the lateral variation of the effective
contact angle will impose a restoring force.

In this paper we use a mesoscopic hydrodynamic model to describe a
nonvolatile fluid on a chemical stripe with such partial stabilization
by realistic (i.e., not step-like) edges. 
We use a numerical scheme based on the thin film approximation, assuming
a sharp liquid-gas interface, partial wetting on both the stripe and
the embedding substrate, small contact angles, and smooth lateral
variations of the disjoining pressure. Analytical estimates can be
obtained within appropriate macroscopic approximations, the validity
of which we discuss in the context of finite size issues.

In accordance with King et al.~\cite{king06}, without drive and
on a homogeneous substrate we find that cylinder-like sessile ridges
are stationary but prone to a Rayleigh-Plateau type pearling instability.
We generalize the analytical macroscopic stability criterion for liquid
ridges on chemical channels with sharp boundaries \cite{brinkmann02,gau99}
to the case of smooth chemical steps. 
A linear stability analysis allows us to account for the occurrence
of large-wavelength pearling, both numerically and analytically. The
presence of a chemical stripe partially stabilizes the ridge with
respect to pearling. The analytical criterion for the stability of
a pinned ridge is in good accordance with the stability domain found
numerically. We also find a quantitative agreement between the growth
rate of long-wavelength pearling obtained from numerical stability
analysis and its corresponding analytical expression obtained by a
perturbative analysis.

External drive applied along the ridge always has an overall stabilizing
effect. 
However, drive never completely stabilizes an unstable liquid ridge
but merely shifts the domain of unstable modes to larger wave lengths.

In the next section \ref{sec:system} we introduce the system we consider
as well as the numerical method we use, along with a discussion of
finite size effects. In Sec.~\ref{sec:thinfilm} we introduce the
dimensionless thin film equation and the parametrization of the chemical
stripe. We analyze the stationary solutions in Sec.~\ref{sec:stationary}
and their linear stability in Sec.~\ref{sec:linstab}. We summarize
and conclude in Sec.~\ref{sec:conclusions}.

\section{Description of the system}

\label{sec:system}

\begin{figure}
\includegraphics[width=1\columnwidth,keepaspectratio,clip]{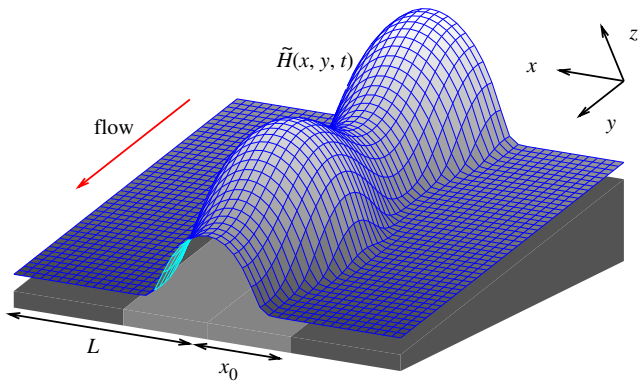}

\caption{\label{cap:setup}Schematic view of the investigated system. The
substrate is topographically flat, invariant in the $y$ direction,
chemically heterogeneous and $2\, L$-periodic in the $x$ direction.
A liquid ridge forms above the lyophilic stripe (grey) of width $2\, x_{0}$
centered on the $y$-axis and a wetting film covers the surrounding
lyophobic regions (black). A body force aligned with the $y$-axis
(caused, e.g., by a substrate tilt) generates flow along the stripe.
The local film thickness at time $t$ is given by $\tilde{H}(x,y,t)$.
Depending on the characteristics of the chemical stripe, the ridge
may be subject to pearling instabilities (exaggerated).}
\end{figure}

As illustrated in Fig.~\ref{cap:setup} we consider a viscous, incompressible,
and nonvolatile fluid on a partially wetting substrate featuring a
straight ``chemical channel'', i.e., a lyophilic stripe with macroscopic
axial extension separating two lyophobic domains. The equilibrium
contact angle is smaller on the channel than on the surrounding, macroscopically
wide, homogeneous substrate. For reasons given below, our analysis
requires a finite spatial extent $2\, L$ in the transverse $x$ direction.
For convenience we assume periodic boundary conditions at $x=\pm L$~.
$L$ is taken to be large enough so that the stripe of width $2\, x_{0}$
can be considered as isolated. In this ``chemical channel'' a
liquid ridge of local thickness can be formed if the amount of liquid
is sufficiently large. The local thickness $\widetilde{H}(x,y,t)$
includes possible variations in the vicinity of the ridge-like stationary
solution.

For such a configuration, the film has an almost uniform thickness
$\Hfilm=\widetilde{H}(\pm L,y,t)$ far from the stripe and one can
define the excess cross sectional area \begin{equation}
\Vex\left(x,y\right)=\int_{-L}^{L}\mathrm{d}x\left[\widetilde{H}(x,y,t)-\Hfilm\right]\label{eq:vex}\end{equation}
of the liquid in the channel. For $y$-invariant stationary solutions
this quantity corresponds to the excess volume of liquid per unit
length on top of the wetting film and thus adequately characterizes
the extra amount of liquid in the system independently of the finite
lateral size $2\, L$.

We consider a finite lateral system size for the following reason.
Given a configuration in which a macroscopic ridge is in stable mechanical
equilibrium with the wetting film on a substrate of finite width,
extending the system size $2\, L$ at constant excess cross-section
eventually leads to an unstable configuration: upon transferring liquid
from the ridge into the wetting film the pressure inside the ridge
increases more than in the wetting film and the ridge will drain into
the film. Therefore the only globally stable configuration on arbitrarily
large substrates is a basically flat wetting film with a slightly
increased thickness above the channel \cite{bauer99d}.

Thus the existence of stable ridges is formally a pure finite size
effect. However, as will be discussed in Sec.~\ref{sec:stationary},
the limit of stability versus pearling 
at a given system size can correspond to a width of the ridge which
is small with respect to the lateral extent of the system. Hence there
is a wide range of configurations for which the quantities relevant
for the dynamics of the ridge are primarily determined by macroscopic
quantities such as the excess cross-section $V_{\mathrm{ex}}$, which
do not depend on the system size.

Within our approach the thin film around the liquid ridge plays an
auxiliary role. It facilitates the mobility of the edges of the ridge,
but contributes negligibly to the dynamics within an appropriately
chosen range of configurations.

A body force aligned with the channel (e.g., gravity if the substrate
is tilted as illustrated in Fig.~\ref{cap:setup} --- the component
normal to the substrate can be neglected) drives the liquid along
the chemical stripe. Our goal is to establish and discuss both analytically
and numerically the conditions of linear stability of a driven flow
in a homogeneously filled channel, in particular with respect to pearling.

Since liquid ridges with contact angles larger than $90^{\circ}$
are unstable even for pinned contact lines, we restrict our analysis
to small contact angles using the thin film approximation \cite{oron97}.
As will be shown below, the translational invariance of the base state
effectively reduces the corresponding boundary value problem to a
set of ordinary differential equations for the base state and to an
eigenvalue problem for ordinary differential equations for the linear
stability analysis. An analytical analysis is possible in the limit
of large wavelengths of the pearling perturbation and without drive.

In the general case, we solve the equations numerically using the
software \noun{Auto2000} \cite{auto}. Within this numerical approach,
instead of looking for a non-trivial solution in a complicated system,
one starts with a simple configuration for which the solution is known.
In the present case the latter is a flat wetting film on a homogeneous
substrate. By gradually incrementing the system parameters towards
non-trivial values, one is able to explore a domain of non-trivial
solutions containing the simple starting point. In the present case,
the most important system parameters are the chemical contrast between
the channel and the embedding substrate, the excess cross-section,
drive, and the wavenumber of the perturbation. 
As for the explicit dependence of the substrate heterogeneity on the
lateral coordinate $x$, treating $x$ as a part of the solution vector
renders the system autonomous as required by \noun{Auto2000}.

\section{Thin film dynamics\label{sec:thinfilm}}

In the limit of small gradients (i.e., long wavelengths), the dynamics
of a thin film of a Newtonian, nonvolatile viscous liquid is well
described by the standard thin film equation for the local film thickness
$\tilde{H}(x,y,t)$ (see, e.g., Ref.~\cite{oron97}). In view of
future purposes, we decompose it into a conservation equation (\ref{eq:conservation}),
an expression (\ref{eq:flow}) for the lateral flow $\mathbf{\tilde{J}}$,
and an equation (\ref{eq:pressure}) for the local pressure $\tilde{P}$
: \begin{subequations} \label{eq:thinfilm} \begin{eqnarray}
\frac{\partial\tilde{H}\left(x,y,t\right)}{\partial t} & = & -\mathbf{\nabla}\cdot\tilde{\mathbf{J}}\left[x,\tilde{H}\left(x,y,t\right)\right]\,,\label{eq:conservation}\\
\tilde{\mathbf{J}}\left(x,\tilde{H}\right) & = & -\frac{Q(\tilde{H})}{\eta}\mathbf{\nabla}\left[\tilde{P}(x,\tilde{H})-\rho gy\right]\,,\label{eq:flow}\\
\tilde{P}\left(x,\tilde{H}\right) & = & -\Pi(x,\tilde{H})-\sigma\mathbf{\nabla}^{2}\tilde{H}\,.\label{eq:pressure}\end{eqnarray}
\end{subequations}
The pressure in Eq.~(\ref{eq:pressure}) is the sum of the Laplace
pressure, which is proportional to the surface tension coefficient
$\sigma$, and the disjoining pressure $\Pi(x,\tilde{H})=-\partial\Phi(x,\tilde{H})/\partial\tilde{H}$.
We choose {[}c.f.~Eq.~(\ref{eq:potential})] the same functional
form for the effective interface potential $\Phi(x,\tilde{H})$ as
frequently used in the context of wetting phenomena (see, e.g., Ref.~\cite{degennes85}
and Refs.~\cite{dietrich88} and \cite{dietrich91} for refined versions).
However, the effective interface potential is an equilibrium concept
and the dynamics in a wetting film a few molecular layers thick is
certainly not given by hydrodynamical equations. In this sense the
repulsive part of $\Pi$ (or $\Phi$) serves the purpose of keeping
the liquid film thickness nonzero even on the lyophobic substrate
and thus allows the three phase contact line to move even in the absence
of hydrodynamic slip at the liquid-substrate interface. Moreover,
as described below, the variation of the contact angle between the
channel and the substrate is encoded in the $x$-dependence of $\Phi$.
The third aspect taken care of by $\Phi$ is to include the influence
of long-ranged dispersion forces on the dynamics in the channel. Thus
we do not expect Eq.~(\ref{eq:thinfilm}) to be an accurate description
of the microscopic dynamics near the contact line and in the wetting
layer; rather, it bears some conceptual similarities with a phase
field equation (a numerical method introduced in the context of crystal
growth, see, e.g., Ref.~\cite{langer80}) for the three phase contact
line. In particular, here it is not necessary to take the location
of the edge into account explicitly. Also, since we do not discuss
wetting transitions within this model, we do not expect a strong dependence
of our results on details of $\Phi$, in particular not on the precise
form of its short-ranged repulsive part and of other subdominant terms.

The additional pressure term $\rho\, g\, y$ in Eq.~(\ref{eq:flow})
models the applied drive with acceleration $g$ of the mass density
$\rho$. The mobility factor $\frac{Q(\tilde{H})}{\eta}=\frac{\tilde{H}^{3}}{3\eta}$
results from the integration of the Poiseuille type velocity profile
over the vertical coordinate. Here we neglect drag by a vapor phase
on the liquid-vapor interface and slip at the substrate \cite{noslip}.

The thickness $\tilde{H}$, the pressure $\tilde{P}$, and the flow
$\tilde{\mathbf{J}}$ depend on the transverse coordinate $x$. Since
the chemical heterogeneity of the substrate is symmetric with respect
to $x=0$ and the driving force is aligned with the $y$ axis, the
stationary configurations exhibit the same symmetry.

Here we focus on the symmetric pearling mode (see Fig.~\ref{cap:setup})
and thus we consider only the interval $x\in[0,L]$. Due to symmetry
and the $2L$-periodicity, odd-order derivatives of the film thickness
with respect to $x$ vanish at $x=0$ and $x=L$, for both the stationary
profile and the perturbation.

In the following the viscosity $\eta$, the surface tension $\sigma$,
and the mass density $\rho$ can be set to 1 by rescaling time, the
lateral coordinates $(x,y)$, and the drive, respectively \cite{drive}.
The remaining degree of freedom, i.e., the scale of the vertical coordinate,
is used to set a certain thickness to 1, which simplifies the expressions
of $\Phi$ and $\Pi$ appearing in our model (see below).

The effective interface potential we use to model partial wetting
is a two-term power law with attraction ($A(x)>0$) at long range
and repulsion ($B(x)>0$) at short range: \begin{equation}
\Phi\left(x,\tilde{H}\right)=-\frac{1}{2}\frac{A(x)}{\widetilde{H}^{2}}+\frac{1}{8}\frac{B(x)}{\widetilde{H}^{8}}.\label{eq:potential}\end{equation}
 The two terms follow from integrating the liquid-liquid and liquid-substrate
Lennard-Jones pair potentials \cite{dietrich91}, assuming a homogenous
substrate. We model a chemically inhomogeneous substrate by effective
amplitudes $A(x)$ and $B(x)$, accounting for a local equilibrium
wetting film thickness $a(x)=\left[\frac{B(x)}{A(x)}\right]^{1/6}$
and an ensuing effective local contact angle \cite{dietrich88} \begin{multline}
\theta_{\mathrm{eq}}(x)=\arccos\left\{ 1+\Phi\left[x,a\left(x\right)\right]/\sigma\right\} \label{eq:theta}\\
=\arccos\left\{ 1-\frac{3A(x)}{8\sigma\left[a(x)\right]^{2}}\right\} \,.\end{multline}

Since in the present study we do not assign a quantitative meaning
to the residual film but use it in order to facilitate contact line
mobility, for numerical convenience we choose $B(x)\propto A(x)$,
so that $a=\mathrm{const}$ is uniform over the \emph{whole} substrate.
According to Eq.~(\ref{eq:theta}) the contact angle contrast is
then provided by the amplitude $A(x)$, which we refer to as the Hamaker
constant, incorporating the commonly used prefactor $\left(6\pi\right)^{-1}$
\cite{degennes85,dietrich88}. Thus, within our model for the chemical
heterogeneity and the thin film approximation, Young's law provides
a local relation between the equilibrium contact angle and the Hamaker
constant: 
\begin{equation}
A(x)=\frac{4}{3}a^{2}\sigma\left[\theq(x)\right]^{2}\,.\label{eq:hamaker_theta}\end{equation}

\begin{figure}
\includegraphics[width=1\columnwidth,keepaspectratio,clip]{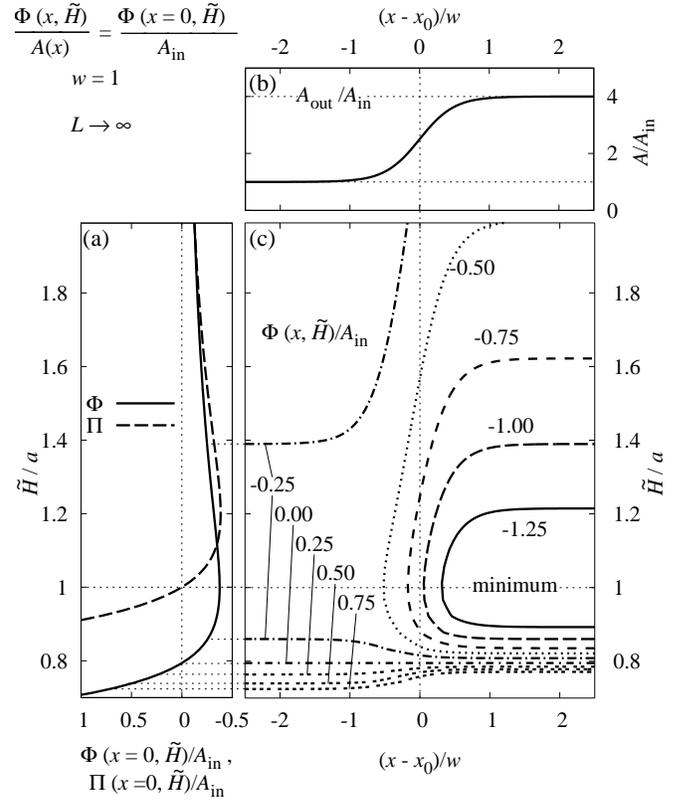}

\caption{\label{cap:potential} Present model of the chemical step of width
$w=1$ with $\Aout=4$ and $\Ain=1$ as used in the numerical analysis.
In the limiting case of infinite lateral extent $2L$, Eq.~(\ref{eq:tanh_arg})
reduces to $f(x)=2\frac{x-x_{0}}{w}$. (a) Effective interface potential
$\Phi$ and disjoining pressure $\Pi$ at the channel center $x=0$
{[}see Eqs.~(\ref{eq:tanh_arg}), (\ref{eq:disjoining_red}), and
(\ref{eq:hamaker_theta_red})]. (b) Laterally varying amplitude $A(x)$
as defined in Eq.~(\ref{eq:hamaker}). (c) Contour plot of the effective
interface potential $\Phi$ near the channel edge. Note that the position
$\tilde{H}=1$ of the minimum of $\Phi$ as a function of $\tilde{H}$
is independent of $x$.}
\end{figure}

In order to describe a chemical channel we choose $A(x)$ to be a
smooth, $2\, L$-periodic function of the transverse coordinate $x$,
symmetric around $x=0$, with well-defined plateau values both inside
the stripe ($\Ain$ for $\left|x\right|\ll x_{0}$) and on the surrounding
substrate ($\Aout$ for $x_{0}\ll\left|x\right|$). At the edges of
the stripe, $A(x)$ varies smoothly over an effective step width $w$.
A chemical channel is thus characterized by the stripe width $2\, x_{0}$,
the step width $w$, and the chemical contrast $\Aout-\Ain$. For
the subsequent numerical analysis we choose the following explicit
functional form for $A(x)$: \begin{subequations} \label{eq:hamaker}
\begin{eqnarray}
A(x) & = & \Ain+\frac{\Aout-\Ain}{2}\left[1-\tanh f(x)\right]\label{eq:hamakerX}\\
f(x) & = & 2L\frac{\cos\frac{\pi x}{L}-\cos\frac{\pi x_{0}}{L}}{\pi w\sin\frac{\pi x_{0}}{L}}.\label{eq:tanh_arg}\end{eqnarray}
\end{subequations}The corresponding structure of the chemical step
near the channel edge is illustrated in Fig.~\ref{cap:potential}.

In order to rescale the thin film equation, we take the residual film
thickness $a$ as the vertical length scale of the problem so that
in these units $\Phi$ attains its minimum at $\tilde{H}=1$. We also
take $A_{\mathrm{in}}$ as a reference Hamaker constant. Equations~(\ref{eq:thinfilm}a-c)
then yield a lateral length scale $\lambda=a^{2}\sqrt{\sigma/A_{\mathrm{in}}}$,
a pressure scale $A_{\mathrm{in}}/a=\sigma a/\lambda^{2}$, a time
scale $\tau=\frac{\eta\,\lambda^{4}}{\sigma\, a^{3}}$, an acceleration
scale $\gamma=\frac{\sigma\, a}{\rho\,\lambda^{3}}$, and a scale
factor for the slopes: $\delta=\frac{a}{\lambda}\ll1$. We note that
these scales, just like $a$ and $A_{\mathrm{in}}$, are arbitrary
(but suitable) and merely provide a consistent way to render the equations
dimensionless with a minimal set of independent system parameters
left. For example, the lateral length scale $\lambda$ is given by
$\lambda=\sqrt{6}\xi_{\parallel}$, where $\xi_{\parallel}=\sqrt{\sigma/\frac{\partial^{2}\Phi}{\partial\tilde{H}^{2}}(a)}$
is the lateral correlation length of the interfacial height-height
correlation function in thermal equilibrium \cite{dietrich88}.

Accordingly, in the following all quantities are dimensionless (i.e.,
effectively, $\sigma$, $\eta$, and $\rho$ are set to 1). In order
to avoid clumsy notations from now on we use the same symbols for
the dimensionless quantities so that 
the dimensionless evolution equation is given by \begin{subequations}
\label{eq:red} \begin{eqnarray}
\frac{\partial\widetilde{H}\left(x,y,t\right)}{\partial t} & = & -\mathbf{\nabla}\cdot\widetilde{\mathbf{J}}\left[x,\widetilde{H}\left(x,y,t\right)\right]\label{eq:conservation_red}\\
\widetilde{\mathbf{J}}\left(x,\widetilde{H}\right) & = & -Q(\widetilde{H})\,\mathbf{\nabla}\left[\widetilde{P}(x,\widetilde{H})-g\, y\right]\label{eq:flow_red}\\
\widetilde{P}\left(x,\widetilde{H}\right) & = & -\Pi(x,\widetilde{H})-\mathbf{\nabla}^{2}\widetilde{H}\label{eq:pressure_red}\end{eqnarray}
 \end{subequations} where $\Pi$ and $Q$ are defined as\begin{equation}
\Pi\left(x,\tilde{H}\right)=-\frac{\partial\Phi}{\partial\widetilde{H}}=-A(x)\left(\frac{1}{\widetilde{H}^{3}}-\frac{1}{\widetilde{H}^{9}}\right)\label{eq:disjoining_red}\end{equation}
and\begin{equation}
Q=\widetilde{H}^{3}/3\,,\label{eq:mobility_red}\end{equation}
while the rescaled Hamaker constant $A(x)$ takes the form: \begin{equation}
A(x)=1+\frac{\Aout/\Ain-1}{2}\left[1-\tanh f(x)\right].\label{eq:hamaker_theta_red}\end{equation}
Since the slopes have been rescaled by $\delta=a/\lambda\ll1$, they
are no longer small. In particular, the rescaled contact angle $\bar{\theta}_{\mathrm{eq}}(x)=\theta_{\mathrm{eq}}/\delta=\frac{1}{2}\sqrt{3A(x)}$
equals $\bar{\theta}_{\mathrm{in}}=\frac{1}{2}\sqrt{3}$ inside the
stripe and $\bar{\theta}_{\mathrm{out}}=\frac{1}{2}\sqrt{3A_{\mathrm{out}}/A_{\mathrm{in}}}$
outside the stripe. In the following we adopt the overbarred notation
for the rescaled contact angle in order to avoid confusion.

\section{Stationary solution\label{sec:stationary}}

For the stationary solution of Eq.~(\ref{eq:red}) there is only
flow along the channel and the system is translationally invariant
in the $y$ direction. The stationary film profile $\tilde{H}_{\mathrm{stat}}=H(x)$
is given by \begin{equation}
-\Pi\left[x,H(x)\right]-H^{\prime\prime}(x)=P,\label{eq:stationary}\end{equation}
which is the same equation as the one characterizing the equilibrium
profile in the absence of flow. Here the pressure $P$ is independent
of $x$, $y$, and t, and it is a free parameter which is determined
by the excess amount $\Vex$ of liquid present in the channel. The
local current is $\mathbf{J}=(0,J_{y}(x),0)$, with $J_{y}(x)=g\, Q\left[H(x)\right]$.

With the disjoining pressure in Eq.~(\ref{eq:disjoining_red}) a
trivial solution of Eq.~(\ref{eq:stationary}) is $H=1$ and $P=0$.
Homogeneous continuation allows us to reach numerically the non-trivial
solutions $H(x)$ by 
continuously varying the parameters of the problem, i.e., $P$ (or
$\Vex$) and the chemical contrast $\Aout-\Ain$. As long as $A_{\mathrm{out}}-A_{\mathrm{in}}=0$~,
parameters such as the effective edge width $w$ do not affect the
trivial solution and thus can be set to desired values before the
continuation.

\begin{figure}
\includegraphics[width=1\columnwidth,keepaspectratio,clip]{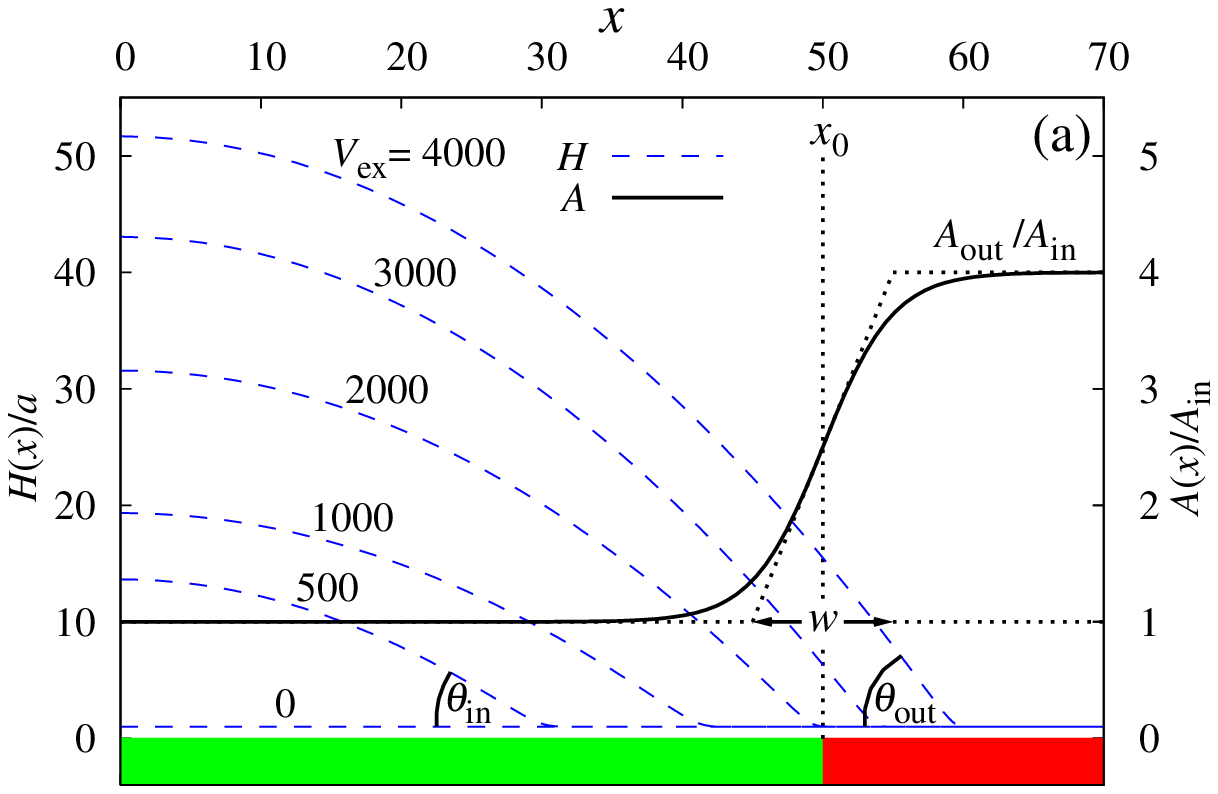}\\
\includegraphics[width=1\columnwidth,keepaspectratio,clip]{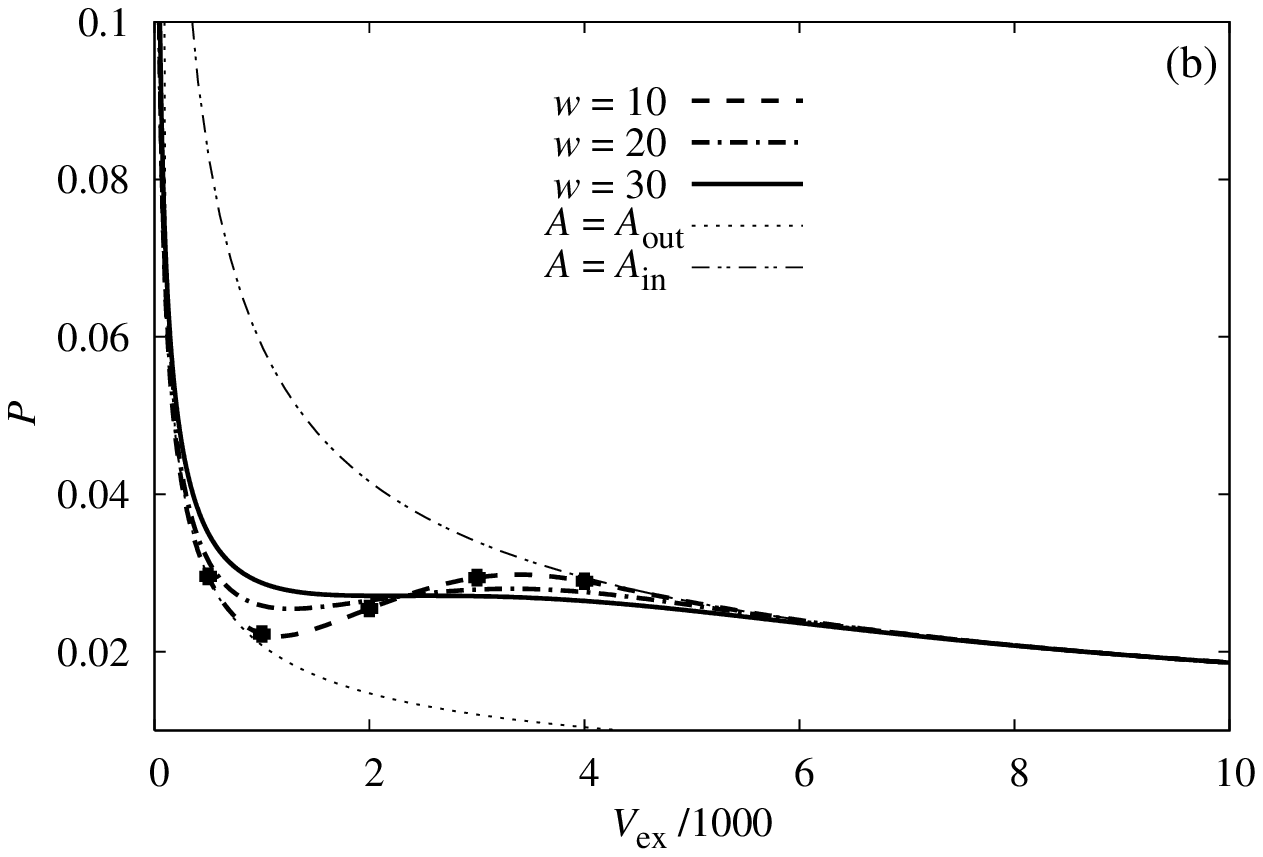}

\caption{\label{cap:pv} (a) Stationary cross section profiles of the liquid
ridge on a chemical channel of half width $x_{0}=50$ and with edges
of effective width $w=10$ for excess cross-sections $\Vex=$~0,
500, 1000, 2000, 3000, and 4000 {[}see Eq.~(\ref{eq:vex})]. Also
shown is the lateral dependence of the Hamaker constant $A$ in units
or $A_{\mathrm{in}}$ and its main features as given by Eq.~(\ref{eq:hamaker}).
$A/A_{\mathrm{in}}=1$ and $A/A_{\mathrm{in}}=A_{\mathrm{out}}/A_{\mathrm{in}}=4$
correspond to $\theta_{\text{in}}=\frac{\sqrt{3}}{2}\delta$ and $\theta_{\text{out}}=\sqrt{3}$$\delta$,
respectively ($\delta=a/\lambda$ is the scale factor for the slopes,
with the lateral scale $\lambda=a^{2}/\sqrt{\sigma/A_{\mathrm{in}}}$).
(b) $\left(P,\Vex\right)$ for the same values of $\Ain$ and $\Aout$
as in (a) for $w=10$ (dashed line, with dots indicating those systems
which the profiles in (a) correspond to), $20$ (dash-dotted line),
and $30$ (solid line)~. The thinner lines correspond to the power
laws $P\propto V_{\mathrm{ex}}^{-1/2}$ given by Eqs.~(\ref{eq:hamaker_theta})
and (\ref{eq:PVmacro}) for ridges resting on homogeneous substrates
corresponding to the outside of the channel (dash-double-dotted) and
to the inside of the channel (dotted). For $w\gtrsim30$~, $P(\Vex)$
is monotonic while for $w\lesssim30$ there is a range of cross-sections
for which $\frac{\mathrm{d}P}{\mathrm{d}\Vex}>0$.
}
\end{figure}

First, we analyze the pressure $P$ as a function of the excess cross-section
$\Vex$ {[}see Eq.~(\ref{eq:vex})] and how it is affected by the
chemical heterogeneity of the substrate 
{[}see Fig.~\ref{cap:pv}(b)]. Each point $\left(\Vex,P\right)$
corresponds to a liquid ridge {[}Fig.~\ref{cap:pv}(a)] centered
around the axis of the chemical channel.
The outer part of the substrate is covered by a film of thickness
close to $1$, which is governed by the disjoining pressure. In the
central region of the ridge, the thickness is large so that due to
the vanishing of $\Pi$ for large $H$ the profile is determined by
the capillary term $H^{\prime\prime}(x)$ in Eq.~(\ref{eq:stationary}).
The edges of the liquid ridge can be located inside the channel 
(i.e., at $\left|x\right|<x_{0}$), be ``pinned'' at the chemical
steps (i.e., at $\left|\left|x\right|-x_{0}\right|\le w$), or ``spill''
onto the surrounding substrate (i.e., at $\left|x\right|>x_{0}$).
In that latter case, the channel will have little effect on the system,
and a spilling ridge has properties similar to those of a ridge on
a homogeneous substrate \emph{without} a chemical channel.

The major features of the $\left(V_{\mathrm{ex}},P\right)$ diagram
in Fig.~\ref{cap:pv}(b) are the following. If the edge of a large
ridge (i.e., apex height $\gg1$) is well inside or well outside the
channel, $P(V_{\mathrm{ex}})$ exhibits a power law $P\propto V_{\mathrm{ex}}^{-1/2}$
with a prefactor which increases with the equilibrium contact angle
$\bar{\theta}_{\mathrm{eq}}$~. In the semi-macroscopic limit the
prefactor is given by Eqs.~(\ref{eq:hamaker_theta}) and, c.f., (\ref{eq:PVmacro}).
For edges pinned at sufficiently sharp chemical steps (i.e., for sufficiently
small $w$) the crossover between the power laws corresponding to
the inner and outer parts of the channel features a change of sign
for the slope $\frac{\mathrm{d}P}{\mathrm{d}V_{\mathrm{ex}}}$.
The latter feature is important for the main focus of our study, because
$\frac{\mathrm{d}P}{\mathrm{d}V_{\mathrm{ex}}}>0$ will turn out as
the criterion of ridge stability.

In the following we first analyze a ridge on a \emph{homogeneous}
substrate, governed by capillarity and resting on a film the thickness
of which is determined by the effective interface potential. From
these semi-macroscopic considerations we derive the power law $P\propto V_{\mathrm{ex}}^{-1/2}$
and discuss features and limitations of the semi-macroscopic model
which will be relevant for the pearling instability to be discussed
in Sec.~\ref{sec:linstab}.

Assuming that the expression for $P$ in Eq.~(\ref{eq:pressure_red})
reduces to $-\mathbf{\nabla}^{2}H$ for the ridge, and to $-\Pi$
for the wetting film, we obtain, for a homogeneous substrate, the
following simple shape for a ridge of half-width $r$: \begin{equation}
H(x)=\left\{ \begin{array}{ccl}
H_{\mathrm{film}}+\frac{P}{2}\left(r^{2}-x^{2}\right) & \text{if} & \left|x\right|<r\,,\\
H_{\mathrm{film}} & \text{if} & r<\left|x\right|\,.\end{array}\right.\label{eq:modelshape}\end{equation}
The wetting film thickness $H_{\mathrm{film}}=1+P/\left(8\bar{\theta}_{\mathrm{eq}}^{2}\right)$
is obtained by solving $-\Pi(H_{\mathrm{film}})=P$ for $H_{\mathrm{film}}\approx1$
and $P\ll1$. The half-width $r$ is determined by the contact angle
according to $\tan(\bar{\theta})=\left|\frac{\partial H}{\partial x}(x=r)\right|\delta$
so that $\bar{\theta}=\theta/\delta\simeq P\, r$. The contact angle
$\bar{\theta}$ of the ridge depends on the pressure $P$ \emph{via}
$H_{\mathrm{film}}$:\begin{equation}
\bar{\theta}=\frac{1}{\delta}\arccos\left[1+\Phi(H_{\mathrm{film}})/\sigma\right]\,.\end{equation}
However, to leading order in $P$ this is simply the equilibrium contact
angle $\bar{\theta}_{\mathrm{eq}}=\frac{1}{\delta}\arccos\left[1+\Phi(a)/\sigma\right]$
given by Eq.~(\ref{eq:hamaker_theta}) and the first correction is
quadratic in $P$.

In order to justify the auxiliary role of the thin film despite the
intrinsic lateral finite-size effect (see Sec.~\ref{sec:system}),
we investigate that regime in which the macroscopic characteristics
of $H(x)$ dominate. For the shape given by Eq.~(\ref{eq:modelshape})
and with $P=\bar{\theta}/r$ one obtains for the cross-section $V=\int_{-L}^{L}H(x)\mathrm{d}x$
and the excess cross-section $V_{\mathrm{ex}}$ the following expressions:
\begin{equation}
V-V_{\mathrm{ex}}=2\, L\,\left(1+\frac{P}{8\,\bar{\theta}^{2}}\right)=2\, L\,\left(1+\frac{1}{8\,\bar{\theta}\, r}\right)\label{eq:VexV}\end{equation}
and\begin{equation}
V_{\mathrm{ex}}=\frac{2}{3}Pr^{3}=\frac{2}{3}\bar{\theta}r^{2}=\frac{2}{3}\bar{\theta}^{3}/P^{2}.\label{eq:PVmacro}\end{equation}
Equation (\ref{eq:PVmacro}) implies that for fixed $\bar{\theta}$
one has $P\propto V_{\mathrm{ex}}^{-1/2}$. In order to ensure that
the film at $x=\pm L$ is indeed flat, one has to choose the system
size $L$ much larger than the ridge width $2\, r$, i.e., $L\gg r$.
Moreover, the scaled thickness of the film outside the ridge should
be close to its equilibrium value $1$, which requires $r\,\bar{\theta}\gg1/8$,
independently of $L$. As long as these two criteria are satisfied,
the excess cross-section $V_{\mathrm{ex}}$ will be, by construction,
independent of the system size $2\, L$.

As we shall show in Sec.~\ref{sec:linstab}, $\frac{\mathrm{d}P}{\mathrm{d}V}$
and $\int Q\,\mathrm{d}x$ are two key quantities which determine
the stability of the liquid ridge.
To a good approximation both should be independent of the system
size.

For the pressure $P$ this means that the derivative with respect
to the total cross-section $V$ should be approximately equal to the
derivative with respect to the excess cross-section $V_{\mathrm{ex}}$.
From Eqs.~(\ref{eq:VexV}) and (\ref{eq:PVmacro}) one obtains 
the ratio of the derivatives with respect to $V_{\mathrm{ex}}$ and
$V$: \begin{multline}
\frac{\mathrm{d}P}{\mathrm{d}V_{\mathrm{ex}}}\,\left(\frac{\mathrm{d}P}{\mathrm{d}V}\right)^{-1}=\frac{\mathrm{d}V}{\mathrm{d}V_{\mathrm{ex}}}\\
=1+2L\frac{\mathrm{d}}{\mathrm{d}V_{\mathrm{ex}}}\left(\frac{P}{8\bar{\theta}^{2}}\right)=1-\frac{3\, L}{16\,\bar{\theta}^{2}\, r^{3}},\label{eq:dPdV}\end{multline}
where the substrate characteristics $L$ and $\bar{\theta}$ are kept
constant. This poses an upper bound on the system size: $3\, L\ll16\,\bar{\theta}^{2}\, r^{3}$,
which quantifies the lateral finite-size effect outlined in Sec. \ref{sec:system}:
if the width $2\, r$ of a ridge is small enough with respect to the
total system size $2L$, the surrounding film can drain liquid from
the ridge while acquiring less additional pressure than the shrunk
ridge. This basic instability of stationary ridges (or drops) on arbitrarily
large substrates is a consequence of the smoothly varying effective
interface potential $\Phi$.

For the shape given in Eq.~(\ref{eq:modelshape}) one can also calculate
the integral over $x$ of the mobility factor $Q(H)=\frac{H^{3}}{3}$.
For sufficiently large ridges (such that $P$ is small and thus the
film thickness is close to $1$, i.e., $r\,\bar{\theta}\gg1/8$) one
obtains \begin{equation}
\int\limits _{-L}^{L}Q\left[H(x)\right]\,\mathrm{d}x=\frac{2\, L}{3}+\frac{2}{3}\,\bar{\theta}r^{2}+\frac{4}{15}\,\bar{\theta}^{2}r^{3}+\frac{4}{105}\bar{\theta}^{3}r^{4}.\label{eq:Qint}\end{equation}
The last term on the right-hand side is the excess mobility integral
which can be written in terms of $V_{\mathrm{ex}}=\frac{2}{3}\bar{\theta}r^{2}$
as \begin{equation}
\left[\int Q\mathrm{d}x\right]_{\text{ex}}=\int\limits _{-L}^{L}Q\left[H(x)-H_{\mathrm{film}}\right]\,\mathrm{d}x=\frac{3\,\bar{\theta}}{35}V_{\mathrm{ex}}^{2}.\label{eq:QVmacro}\end{equation}
The second and the third term on the right hand side of Eq.~(\ref{eq:Qint})
are negligible: the respective ratios with the excess mobility integral
are $\propto(\bar{\theta}\, r)^{-2}$ and $\propto(\bar{\theta}\, r)^{-1}$,
respectively, and hence small in the limit of large $r$ considered
here (see above). For $\bar{\theta}^{3}r^{4}\gg\frac{35}{2}L$ the
excess mobility integral dominates the contribution from the film.
A similar argument can be formulated in the presence of slippage at
the substrate, i.e., for a more general expression of $Q$.

The lower bound on $L$ is thus $\propto r$ (so that the liquid ridge
does not interfere with its periodic images), while the upper bounds
are $\propto r^{3}$ {[}see the text after Eq.~(\ref{eq:dPdV})]
and $\propto r^{4}$ (see above). This implies that for every large
enough $r$ one can find a range of system sizes $L$ within which
the key quantities of the linear stability (as discussed in the following
section) are independent of $L$ and of the macroscopic ridge cross-section
$V_{\mathrm{ex}}$.

\section{Linear stability analysis\label{sec:linstab}}

In order to assess the stability of a liquid ridge on a chemical channel,
we consider the time evolution of small perturbations of the stationary
film thickness as well as the corresponding small perturbations of
the pressure and the flows. Since the base state is translationally
invariant in the $y$ direction, we consider perturbations with the
form of plane waves:\begin{subequations} \label{eq:complex} \begin{eqnarray}
\widetilde{H}(x,y,t) & = & H(x)+\varepsilon\, h(x)\, e^{\omega\, t-i\, k\, y},\label{eq:thickness_complex}\\
\widetilde{P}(x,y,t) & = & P+\varepsilon\, p(x)\, e^{\omega\, t-i\, k\, y},\label{eq:pressure_complex}\\
\widetilde{J}_{x}(x,y,t) & = & 0+\varepsilon\, j_{x}(x)\, e^{\omega\, t-i\, k\, y},\label{eq:flow_complex}\\
\widetilde{J}_{y}(x,y,t) & = & J_{y}(x)+\varepsilon\, j_{y}(x)\, e^{\omega\, t-i\, k\, y},\end{eqnarray}
\end{subequations}with $\varepsilon\ll1$~. 
Insertion into Eqs.~(\ref{eq:red}a-c) and expansion to first order
in $\varepsilon$ leads to the following linear eigenvalue problem
for the complex growth rate $\omega(k)$:\begin{subequations}\label{eq:Lin}
\begin{eqnarray}
\omega(k)\, h & = & -\frac{\mathrm{d}j_{x}}{\mathrm{d}x}-k^{2}\, Q\, p+i\, k\, g\,\left.\frac{\mathrm{d}Q}{\mathrm{d}\widetilde{H}}\right|_{H}h\,,\label{eq:conservationLin}\\
j_{x} & = & -Q\,\frac{\mathrm{d}p}{\mathrm{d}x}\,,\label{eq:flowLin}\\
p & = & \left(k^{2}-\left.\frac{\partial\Pi}{\partial\widetilde{H}}\right|_{H}\right)\, h-\frac{\mathrm{d}h'}{\mathrm{d}x}\,,\label{eq:pressureLin}\\
h' & = & \frac{\mathrm{d}h}{\mathrm{d}x}\,.\label{eq:slopeLin}\end{eqnarray}
 \end{subequations}

In order to solve this problem numerically 
by continuation from a simple configuration, one must integrate Eq.~(\ref{eq:stationary})
for the stationary profile and the linearized equations (\ref{eq:Lin})
together. As in Sec.~(\ref{sec:stationary}), we render the system
autonomous by introducing $s=\frac{x}{L}$ and converting $x$ into
a component of the solution vector. This leads to the following first-order
system of non-linear equations: \begin{equation}
\frac{\mathrm{d}}{\mathrm{d}s}\left(\begin{array}{c}
x\\
H\\
H'\\
h\\
h'\\
p\\
j_{x}\end{array}\right)=L\,\left(\begin{array}{c}
1\\
H'\\
-\Pi-P\\
h'\\
\left(k^{2}-\frac{\partial\Pi}{\partial H}\right)h-p\\
-j_{x}/Q\\
-\left(\omega-ikg\frac{\mathrm{d}Q}{\mathrm{d}H}\right)h-k^{2}Qp\end{array}\right).\label{eq:system}\end{equation}
We note that $h$, $h'$, $p$, and $j_{x}$ are in general complex
valued functions of $s$. Only for $g=0$ there are real solutions.

The spectrum is semi-discrete. The wavevector $k$ along the $y$
direction is continuous, while the solutions of Eq.~(\ref{eq:system})
for a given $k$ are generated by a discrete family of modes due to
the finite lateral extent $L$. We are primarily interested in the
varicose or pearling mode, which is obtained by continuation from
the fundamental symmetric mode, which in the simple case of a homogeneous
substrate covered by a film of thickness $H=1$ corresponds to perturbations
in the $y$ direction only: \begin{eqnarray}
h & = & 1\,,\label{eq:thicknessLin0}\\
j_{x} & = & 0\,,\label{eq:flowLin0}\\
p & = & k^{2}+6A\,,\label{eq:pressureLin0}\\
\omega(k) & = & ikg-k^{2}\left(k^{2}+6A\right)\,.\label{eq:growthLin0}\end{eqnarray}
Since $A=\frac{4}{3}\,\bar{\theta}^{2}>0$, $\mathrm{Re}\,\omega=-k^{2}\left(k^{2}+6A\right)<0$
and for any $k\neq0$ the fundamental symmetric mode is unconditionally
stable on a homogeneous substrate covered by a flat residual film.
It is only in homogeneous situations that this mode (i.e., a plane
wave along the $y$-direction) has no nodes along the $x$-direction
{[}see, c.f., Eq.~(\ref{eq:fundamental})].

\subsection{Systems without drive: $g=0$}

\label{sec:nodrive}The Rayleigh-Plateau instability occurs for long
wavelengths, i.e., $k\rightarrow0$. For large $k$, however, surface
tension will stabilize all perturbations and we expect $\omega(k)\sim-k^{4}$
in this limit. In order to assess the stability of a liquid ridge,
we therefore focus on the limit $k\to0$. Since the mode with $k=0$
is marginally stable due to volume conservation, we have to infer
the stability or instability of a ridge from the eigenvalue for small
nonzero values of $k$.

\begin{figure}
\includegraphics[width=1\columnwidth,keepaspectratio,clip]{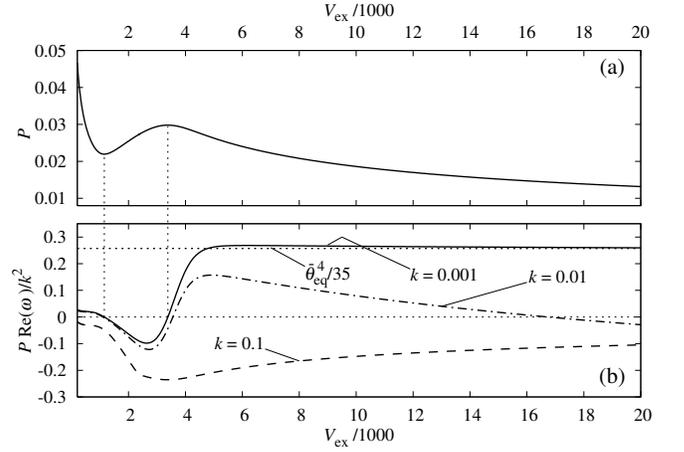}

\caption{\label{cap:nodrive1} Part (a) shows the $\left(P,\Vex\right)$ diagram
of the stationary solutions for a given substrate heterogeneity ($\Ain=1$,
$\Aout=4$, $w=10$, $x_{0}=50$; compare Fig.~\ref{cap:pv}). In
this case the chemical steps are sharp enough for a change of sign
of $\frac{\mathrm{d}P}{\mathrm{d}\Vex}$ in a finite interval of excess
cross sections $V_{\mathrm{ex}}$. As function of $V_{\mathrm{ex}}$
part (b) shows the growth rate $\mathrm{Re}\,\omega(k;\Vex)$ {[}divided
by $k^{2}$ and multiplied by $P(\Vex)$] of long-wavelength varicose
modes for several small values of $k$. The stability window of the
varicose mode corresponds to the range where $\frac{\mathrm{d}P}{\mathrm{d}\Vex}>0$.
Note that for small wave numbers $k$ and large $\Vex$, $P\,\omega(\Vex,k)/k^{2}$
approaches a plateau {[}with a value predicted by Eq.~(\ref{eq:omegahomo})
to be $\frac{9}{35}=0.26$; for large $V_{\mathrm{ex}}$ one has $\bar{\theta}_{\mathrm{eq}}^{2}=3A_{\mathrm{out}}/4=3$,
which is consistent with the observed value]. On the present scales
the curves for $k\lesssim10^{-3}$ would be barely distinguishable
from the one for $k=10^{-3}$. For $k=0.01$ a typical stabilization
can be seen for $V_{\mathrm{ex}}\gtrsim17000$, as the wavelength
corresponding to $k$ becomes short with respect to the ridge width.
For $k=0.1$ a finite-size effect is observed for $V_{\mathrm{ex}}\gtrsim2500$:
in this case the mode does no longer correspond to a bulging of the
ridge but to system-wide perturbations of the film, which are stable.
}
\end{figure}

Figure~\ref{cap:nodrive1} shows the pressure $P$ for a chemical
channel with $w=10$ {[}compare Fig.~\ref{cap:pv}(b)] and the (purely
real) growth rate $\omega$ of long-wavelength perturbations {[}determined
numerically by solving Eq.~(\ref{eq:system})] as a function of the
excess cross-section $V_{\mathrm{ex}}$. We emphasize two features
of this figure.

First, the juxtaposition of the two graphs clearly suggests $\frac{\mathrm{d}P}{\mathrm{d}V_{\mathrm{ex}}}>0$
as a stability criterion for large-wavelength deformations of the
ridge. All long-wavelength modes possess a stability domain {[}$\omega(k;\Vex)<0$],
which includes the range of $V_{\mathrm{ex}}$ for which $\frac{\mathrm{d}P}{\mathrm{d}V_{\mathrm{ex}}}>0$.
These are also the cross-sections for which the edges of the ridge
are pinned at the channel edge (see Fig.~\ref{cap:pv}). 

Second, for values of $V_{\mathrm{ex}}$ for which the edges of the
base state of the liquid ridge are well outside the chemical channel,
$\omega(V_{\mathrm{ex}},k)\, P(V_{\mathrm{ex}})/k^{2}$ converges
to a constant independent of $V_{\mathrm{ex}}$ for $k\to0$. 
This is characteristic for ridges resting on homogeneous substrates,
and we shall derive a corresponding analytic expression below. 

The first observation has a simple explanation. 
A long wavelength perturbation corresponds to a periodic arrangement
of liquid bumps separated by thinned regions. For small $k$, the
transition regions between the bumps and the thinned regions can be
ignored and the pressure in such a bump is approximately given by
the pressure in a homogeneous ridge of corresponding cross section.
If $\frac{\mathrm{d}P}{\mathrm{d}V}>0$, the pressure in the thicker
part will be larger than in the thinner part and the liquid will flow
from the bump to the thin part, leveling the perturbation. If, on
the other hand, $\frac{\mathrm{d}P}{\mathrm{d}V}<0$, the pressure
in the bump will be smaller than in the thin part, the bump will be
inflated (for the same reason as a small soap bubble inflates a larger
one), and the perturbation will grow.

In the following we shall confirm the phenomenological stability criterion
$\frac{\mathrm{d}P}{\mathrm{d}V}>0$ by a perturbation analysis for
small $k$ of the eigenvalue problem formulated in Eq.~(\ref{eq:Lin}).
To this end, for $g=0$ we write Eq.~(\ref{eq:Lin}) as a differential
equation of fourth order for $h(x;k)$: \begin{equation}
\omega(k)\, h(x;k)=\left(\hat{F}+\hat{K}\right)\, h(x;k)\label{eq:eigen}\end{equation}
with the $k$-independent linear operator \begin{equation}
\hat{F}=-\frac{\partial}{\partial x}Q(H)\frac{\partial}{\partial x}\left(\frac{\partial\Pi}{\partial H}+\frac{\partial^{2}}{\partial x^{2}}\right)\label{eq:timeop}\end{equation}
and the $k$-dependent linear operator \begin{equation}
\hat{K}=k^{2}\, Q(H)\,\left(\frac{\partial\Pi}{\partial H}+\frac{\partial}{\partial x^{2}}-k^{2}\right)+k^{2}\frac{\partial}{\partial x}\left[Q(H)\frac{\partial}{\partial x}\right].\label{eq:perturbationop}\end{equation}
Both $\hat{F}$ and $\hat{K}$ act on $h(x;k)$ via multiplication
followed by $\frac{\partial}{\partial x}$. With the usual scalar
product $\langle\phi|\psi\rangle=\frac{1}{2\, L}\int_{-L}^{L}\bar{\phi}(x)\,\psi(x)\,\mathrm{d}x$
in the space of $2\, L$-periodic complex valued functions, where
the overbar indicates complex conjugation, the adjoint operator to
$\hat{F}$ is \begin{equation}
\hat{F}^{\dagger}=-\left(\frac{\partial\Pi}{\partial H}+\frac{\partial^{2}}{\partial x^{2}}\right)\,\frac{\partial}{\partial x}\left[Q(H)\frac{\partial}{\partial x}\right]\,.\label{eq:adjoint}\end{equation}
By construction, $\hat{F}$ and $\hat{F}^{\dagger}$ have the same
discrete spectrum of eigenvalues $\omega_{n}(k=0)$ and the eigenfunctions
$f_{n}(x)$ (of $\hat{F}$) and $f_{n}^{*}(x)$ (of $\hat{F}^{\dagger}$,
also called left eigenfunctions of $\hat{F}$) to different eigenvalues
are orthogonal, i.e., $\langle f_{n}^{*}|f_{m}\rangle=\delta_{nm}$.
(Recall that $\bar{f}_{n}$, not $f_{n}^{*}$, is the complex conjugate
to $f_{n}$.) Since both operators commute with the parity operator,
in the following we can restrict our analysis to modes symmetric with
respect to $x=0$. As we already observed for the numerical solution,
the mode relevant for the pearling instability is the fundamental
symmetric mode. By differentiating Eq.~(\ref{eq:stationary}) with
respect to $P$ one can show that $\frac{\mathrm{d}H}{\mathrm{d}P}(x)=\frac{\mathrm{d}H}{\mathrm{d}V}(x)/\frac{\mathrm{d}P}{\mathrm{d}V}$
is an eigenfunction of $\hat{F}$ corresponding to the eigenvalue
$\omega_{0}(k=0)=0$. The normalized set of fundamental modes is then
\begin{equation}
f_{0}(x)=2\, L\,\frac{\mathrm{d}H(x)}{\mathrm{d}V}\quad\text{and}\quad f_{0}^{*}(x)=1\,.\label{eq:fundamental}\end{equation}
Although $f_{0}(x)$ is the fundamental mode, in the interval $[0,L]$
it can have a (single) zero at the edge of the ridge. On the ridge
$\frac{\mathrm{d}H(x)}{\mathrm{d}V}$ is positive, whereas on the
surrounding flat film it is negative if $\frac{\mathrm{d}P}{\mathrm{d}V}<0$.
One can show that $\langle f_{0}^{*}|f_{0}\rangle=1$ by swapping
the integration with respect to $x$ and the differentiation with
respect to $V$. The model shape given by Eq.~(\ref{eq:modelshape}),
where $H_{\mathrm{film}}=1+P/\left(8\,\bar{\theta}_{\mathrm{eq}}^{2}\right)$,
provides an instructive illustration.

For small $k$, we assume that the fundamental eigenmode $h_{0}\left(x;k\right)$
differs only slightly from $f_{0}(x)$ and we can expand it (up to
a normalization constant which we do not need to consider) in terms
of eigenfunctions of $\hat{F}$: \begin{equation}
h_{0}(x;k)=f_{0}(x)+\sum\limits _{n>0}a_{n}(k)\, f_{n}(x)\,,\label{eq:expansion}\end{equation}
with expansion coefficients $a_{n}(k)$ of order $k$. One obtains
the lowest order correction to $\omega_{0}(k)$ by inserting $h_{0}(x;k)$
from Eq.~(\ref{eq:expansion}) into Eq.~(\ref{eq:eigen}) and projecting
the result onto $f_{0}^{*}$, keeping only terms up to order $k^{2}$.
With $\langle f_{0}^{*}|f_{0}\rangle=1$ one finds \begin{multline}
\omega_{0}(k)=\omega_{0}(0)+k^{2}\,\left\langle f_{0}^{*}\left|Q(H)\left(\frac{\partial\Pi}{\partial H}+\frac{\partial^{2}}{\partial x^{2}}\right.\right)f_{0}\right\rangle \\
+k^{2}\,\left\langle f_{0}^{*}\left|\frac{\partial}{\partial x}\left[Q(H)\frac{\partial}{\partial x}f_{0}\right]\right.\right\rangle .\end{multline}
The nature of $f_{0}$ implies $\omega_{0}(0)=0$ as detailed previously.
The integrand in the scalar product in the last term is a total derivative
so that for a $2\, L$-periodic function this term vanishes. For the
first scalar product on the right hand side the integrand reduces
to $Q(H)\,\frac{\mathrm{d}P}{\mathrm{d}V}$ which implies \begin{equation}
\omega_{0}(k)=-k^{2}\,\frac{\mathrm{d}P}{\mathrm{d}V}\,\int\limits _{-L}^{L}Q\left[H(x)\right]\,\mathrm{d}x\,.\label{eq:growth_analytic}\end{equation}
We note that Eq.~(\ref{eq:growth_analytic}) is valid for a general
expression of the mobility factor $Q(H)$, not only for the no-slip
case studied numerically. In general the velocity field is such that
$Q(H)>0$, so that the integral in Eq.~(\ref{eq:growth_analytic})
is positive. This confirms the observation made for the numerical
solution and supports the heuristic argument given at the beginning
of this subsection, i.e., that the stationary ridge is stable for
$\frac{\mathrm{d}P}{\mathrm{d}V}>0$. A similar discussion can be
carried out for a more general mobility factor which could depend
explicitly on $x$.

For the macroscopic ridge described by Eq.~(\ref{eq:modelshape})
one has $\frac{\mathrm{d}P}{\mathrm{d}V}\approx\frac{\mathrm{d}P}{\mathrm{d}V_{\mathrm{ex}}}=\frac{\mathrm{d}P}{\mathrm{d}r}/\frac{\mathrm{d}V_{\mathrm{ex}}}{\mathrm{d}r}$.
The generalization of Eq.~(\ref{eq:PVmacro}) to heterogeneous substrates
by introducing an effective local contact angle $\bar{\theta}_{\mathrm{eq}}(x)$
leads, at the ridge edge ($x=r$), to $P=\bar{\theta}_{\mathrm{eq}}(r)/r$
and $V_{\mathrm{ex}}=\frac{2}{3}\, r^{2}\,\bar{\theta}_{\mathrm{eq}}(r)$.
Since the equilibrium contact angle inside the channel is smaller
than outside one has $\mathrm{d}\bar{\theta}_{\mathrm{eq}}(r)/\mathrm{d}r>0$
and therefore $\mathrm{d}V_{\mathrm{ex}}/\mathrm{d}r>0$. However,
the pressure as a function of the half ridge width $r$ can be nonmonotonous.
From $\frac{\mathrm{d}P}{\mathrm{d}r}=\left[\mathrm{d}\bar{\theta}_{\mathrm{eq}}(r)/\mathrm{d}r-\bar{\theta}_{\mathrm{eq}}(r)/r\right]/r$
one obtains as a macroscopic stability criterion \begin{equation}
\frac{1}{\bar{\theta}_{\mathrm{eq}}(r)}\,\frac{\mathrm{d}\bar{\theta}_{\mathrm{eq}}(r)}{\mathrm{d}r}=\frac{\mathrm{d}\ln\bar{\theta}_{\mathrm{eq}}(r)}{\mathrm{d}r}>\frac{1}{r}.\end{equation}

For a homogeneous substrate $\omega_{0}(k)$ can be determined in
the macroscopic limit discussed in Sec.~\ref{sec:stationary}. For
a given contact angle $\bar{\theta}_{\mathrm{eq}}$ Eqs.~(\ref{eq:PVmacro})
and (\ref{eq:QVmacro}) yield \begin{equation}
\omega_{0}(k)=k^{2}\,\frac{1}{7}\sqrt{\frac{3\,\bar{\theta}_{\mathrm{eq}}^{5}}{2}}\,\sqrt{\Vex}=k^{2}\,\frac{\bar{\theta}_{\mathrm{eq}}^{4}}{35}\,\frac{1}{P}\label{eq:omegahomo}\end{equation}
as a function of $\Vex$ or $P$, respectively. These relations provide
an understanding for the numerical observation in Fig.~\ref{cap:nodrive1}
that for small $k$ the growth rate rescaled by $k^{2}/P$ approaches
a constant as $V_{\mathrm{ex}}\to\infty$, i.e., in the limit of a
homogeneous substrate.

\subsection{Systems with drive: $g\neq0$}

\label{sec:drive}

\begin{figure}
\includegraphics[width=1\linewidth,keepaspectratio,clip]{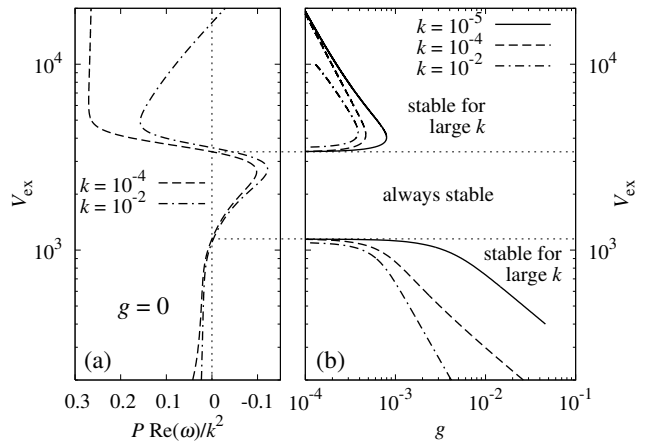}

\caption{\label{cap:drive1}For reference purposes part (a) corresponds to
the rotated Fig.~\ref{cap:nodrive1}(b), obtained for the case without
drive, with the curve for $k=0.1$ removed. Part (b) traces the two
zeros of $\Real\,\omega(k)$ as a function of the drive $g$ by showing
the isolines of $\Real\,\omega(k;\Vex,g)=0$ for selected small values
of $k$. Ridges with cross-sections between the horizontal lines are
stable. Other ridges are unstable with respect to pearling, with the
marginally stable wavelength depending on the drive. The channel parameters
are the same as in Fig.~\protect\ref{cap:nodrive1}.}

\end{figure}

\begin{figure}
\includegraphics[clip,width=1\linewidth,keepaspectratio]{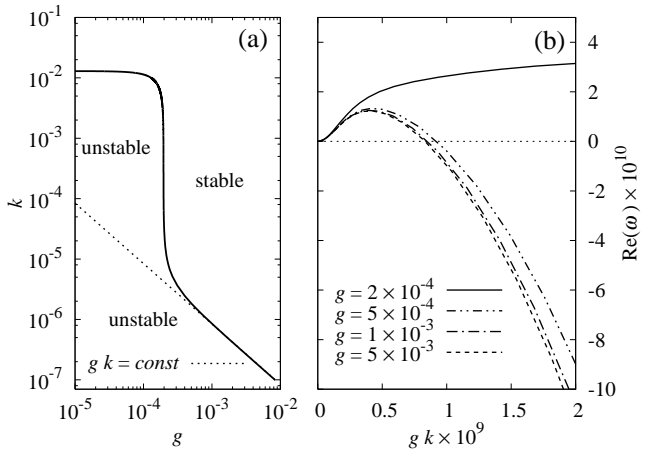}

\caption{\label{cap:drive2}(a) Curve of marginal stability, $\Real\,\omega(k;g,\Vex)=0$,
in the $g$-$k$-plane (full line) for $\Vex=10^{4}$ (i.e., outside
the stability region for $g=0$, see Fig.~\ref{cap:drive1}) and
the same channel parameters as in Figs.~\ref{cap:nodrive1} and \ref{cap:drive1}.
For large $g$, the drive required to stabilize modes with large wavelengths
is proportional to $1/k$. The straight dashed line is a fit $k\propto1/g$
to the curve for $g>10^{3}$. (b) The growth rate $\Real\,\omega(k;g,\Vex)$
as function of $g\, k$ for the same ridge and several values $g$.
For large $g$ the curves approach a limiting curve $\omega_{\infty}(g\, k;\Vex)$,
which depends on $g$ and $k$ only via the product $g\, k$.
}
\end{figure}

Figure~\ref{cap:drive1}(b) shows the isolines $\Real\,\omega(k;g,\Vex)=0$
in the $g$-$\Vex$-plane for several values of $k$. For $g=0$,
the corresponding values of $V_{\mathrm{ex}}$ are the zeros of the
curves in Fig.~\ref{cap:nodrive1}(b), repeated for convenience in
Fig.~\ref{cap:drive1}(a). For small values of $k>0$, as function
of $g$ the isolines follow the two horizontal lines, which mark the
lower and upper cross-sections for which $\frac{\mathrm{d}P}{\mathrm{d}\Vex}=0$,
up to larger values of $g$ before they turn away from these lines.
However, they never penetrate the range of cross-sections between
the two horizontal lines. Thus ridges in this range of excess cross-sections
remain stable also under drive. For ridges with $V_{\mathrm{ex}}$
outside of this range Fig.~\ref{cap:drive1}(b) leads to the following
conjectures. For any $V_{\mathrm{ex}}$, modes with a given $k$ are
stabilized by a large enough drive $g>g_{c}(k,V_{\mathrm{ex}})$ (see
below). On the other hand, for every ridge and every finite value
of $g$ there is a sufficiently small wave number $k_{c}(g,V_{\mathrm{ex}})$
such that modes with smaller wave numbers $k<k_{c}(g,V_{\mathrm{ex}})$
are unstable. Therefore drive cannot stabilize a liquid ridge versus
pearling as such, but it shifts the critical wavelength for the onset
of instability to larger values. Hence with driven flow we expect
the appearance of larger pearls which also emerge further apart from
each other.

For $g\ne0$ one needs the full eigenvalue spectrum $\omega_{n}(k)$,
for all $n\geq0$, in order to compute the lowest order correction
to $\omega_{0}(0)$ for small $k$ in the perturbation analysis presented
in Subsec.~\ref{sec:nodrive}. However, for large $g$ and small
$k$ we expect that the last term in Eq.~(\ref{eq:conservationLin}),
i.e., the term proportional to $g$, dominates the eigenvalue problem,
which turns $\omega(k,g)$ into a function of $g\, k$. We have confirmed
this expectation numerically. Figure~\ref{cap:drive2}(a) shows the
stability boundary for a given ridge cross-section $V_{\mathrm{ex}}$
in the $g$-$k$-plane. For large $g$ the critical wave number $k_{c}(g,V_{\mathrm{ex}})$,
for which $\mathrm{Re}\,\omega\left[k_{c}(g,V_{\mathrm{ex}});g,V_{\mathrm{ex}}\right]=0$,
is indeed proportional to $1/g$~. Modes with smaller $k$ (or longer
wavelength) require a larger drive to be stabilized. However, for
every drive $g$ there are unstable modes. For a given cross-section
$V_{\mathrm{ex}}$ and increasing values of $g$ Fig.~\ref{cap:drive2}(b)
shows the growth rate $\mathrm{Re}\,\omega(k;g,V_{\mathrm{ex}})$
as a function of $g\, k$ for increasing values of $g$. As expected
the curves converge to a limiting curve $\omega_{\infty}(g\, k;V_{\mathrm{ex}})$.
Interestingly, $\omega_{\infty}(g\, k;V_{\mathrm{ex}})\sim(g\, k)^{2}$
for large as well as for small values of $g\, k$, but with different
prefactors (positive for $gk\lesssim10^{-10}$, negative for $gk\gtrsim10^{-9}$).

\section{Summary and conclusions}

\label{sec:conclusions}As illustrated in Fig.~\ref{cap:setup},
we have used the lubrication approximation in order to analyze the
stability of nonvolatile liquid ridges versus pearling in the case
of driven flow along a chemical stripe with smooth edges (see Fig.~\ref{cap:potential}).
Such ridges can be stable versus pearling even though their contact
lines are not completely pinned (see Fig.~\ref{cap:pv}). In an analytic
perturbation analysis for small wave numbers $k$ as well as numerically
(see Fig.~\ref{cap:nodrive1}) we have confirmed $\frac{\mathrm{d}P}{\mathrm{d}V}>0$
as the corresponding stability criterion, i.e., a ridge is stable
versus pearling if the pressure $P$ in the ridge increases with its
cross-section $V$. If the ridge is guided by a chemical channel with
smooth edges, which can be characterized by a laterally varying effective
contact angle $\theta_{\mathrm{eq}}(x)$, we find the stability criterion
$\frac{\mathrm{d}\ln\theta_{\mathrm{eq}}(x)}{\mathrm{d}x}>\frac{1}{x}$
where $x$ is the lateral distance from the center of the stripe.
This criterion for the chemical design of the stripe also holds if
a body force aligned with the channel drives the liquid.

We find a stabilizing effect of drive (see Fig.~\ref{cap:drive1}).
Stable modes remain stable under drive but the critical wavenumbers
$k_{c}$ (such that modes with $k<k_{c}$ are unstable) decrease with
drive. Therefore, whenever the ridge is indeed subject to the pearling
instability, we expect the size of the emerging pearls and the distance
between them to increase with drive. For any finite drive, there is
a nonzero range of long wavelengths for which the modes are unstable.
Hence the pearling instability cannot be suppressed by flow. These
findings are in agreement with the results in Ref.~\cite{koplik06a}
for well-filled ridges with large contact angles $\ge90^{\circ}$
and a fixed contact line. However, in contrast to the case of well-filled
ridges, in the thin film limit the maximum of the growth rate $\Real\,\omega$
of unstable modes does not increase with drive (see Fig.~\ref{cap:drive2}).

We thank M. Brinkmann for fruitful discussions. M. Rauscher acknowledges
financial support from the priority program SPP~1164 ``Micro and
Nano Fluidics'' of the Deutsche Forschungsgemeinschaft under grant
number RA~1061/2-1.


\begin{thebibliography}{10}
\bibitem{noslip}In practice, the velocity profile is affected
by the heterogeneous chemical properties of the substrate, in particular
via the Navier slip length. This would lead to an explicit dependence
of the mobility factor $Q$ on the transverse coordinate $x$. However,
in our study we assume, unless mentioned otherwise, that the no-slip
condition is fulfilled on the whole substrate, leading to the simple
expression $Q=\tilde{H}^{3}/3$, which depends on $x$ only via $\tilde{H}$.
Some analytical results still hold for general and explicitly $x$-dependent
expressions for $Q(x;\tilde{H}(x))$.
\bibitem{drive}Here the picture of gravitational drive {[}Eq.~(\ref{eq:flow})]
is used as an example, but the present study covers any kind of driving
force acting on the bulk.
\end{thebibliography}
\end{document}